\documentstyle[11pt,newpasp,twoside,epsf]{article}
\def\edcomment#1{\iffalse\marginpar{\raggedright\sl#1\/}\else\relax\fi}
\reversemarginpar

\begin{document}
\title{Crowding in the GAIA spectrograph focal plane}
 \author{Toma\v{z} Zwitter}
\affil{University of Ljubljana, Dept.\ of Physics, Jadranska 19,
1000 Ljubljana, Slovenia, tomaz.zwitter@fmf.uni-lj.si}
\author{Arne Henden}
\affil{Universities Space Research Association/U.S.\ Naval Observatory,
Flagstaff AZ 86002-1149, USA, aah@nofs.navy.mil}

\begin{abstract}
Superpositions of stellar tracings are present in every slitless 
spectrograph. The probability for such overlaps in the GAIA 
RVS spectrograph focal plane is estimated using photometric 
observations of 66 stellar fields, mostly close to the Galactic 
plane. It is shown that overlaps of bright stars ($V<17$) are 
common near the Galactic plane, and no spectrum is free
from superpositions of faint star tracings.
Most overlappers are of spectral type K. 
\end{abstract}

\section{Introduction}

GAIA RVS is a slitless spectrograph, so some degree of crowding
due to spectral tracing overlaps in the focal plane is to be 
expected. Overlapping spectra increase the effective background 
signal and make it highly ununiform. In an acompanying paper 
(Zwitter 2002) we show that superpositions of spectral 
lines of background stars can be removed by careful modelling. 
Here we use photometry of actual stellar fields in 66 directions
in the Galaxy to assess the probability for an overlap, as well as 
the typical spectral type of overlappers. 

\section{Observations}

We use B, V, and I$_{\mathrm C}$ photometry of 66 fields around a sample of 
symbiotic stars (Henden \&\ Munari 2000, 2001). These fields
were observed with the USNO Flagstaff Station, 1.0m telescope
and two CCD detectors.  The exposures were designed to
permit accurate photometry of the symbiotic star, and so
were of varying depth depending on the brightness of the
target star.  However, all fields are relatively
complete to $V \sim 17$ with stellar 
detections reaching $V=20.5$.  Only photometric, good seeing
nights were used, with extinction and transformation coefficients
determined from nightly measures of Landolt standard stars.
Aperture photometry was used, with a minimum of three observations
on separate photometric nights per field.  Typical zeropoint
errors are around $0.01$mag.  More complete description of
the observations and techniques can be found in the
referenced papers.  These fields can be used to assess stellar 
density and distribution of spectral types down to the GAIA faintness 
limit. A total of 75959 stars were analyzed. 

\begin{figure}
\plotfiddle{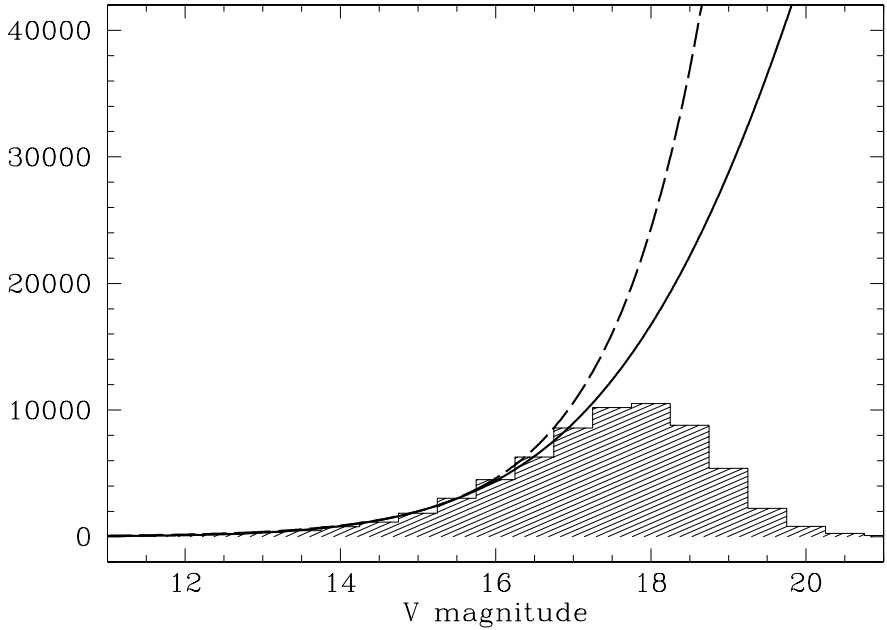}{7.9cm}{270}{48}{48}{-190}{260}
\caption{Luminosity function from 66 stellar fields with 
76000 stars. Histogram represents magnitude distribution of 
V-magnitude measurements, solid line is the assumed true distribution
(eq.\ 1), while dashed line is a potential fit ($dN/dV \propto 2.3^V$) 
to the bright stars ($V<16$).}
\end{figure}

The luminosity function for $V>15$ is frequently described by a 
potential law. Here we adopt a heuristic law
\begin{equation}
dN/dV \propto 2.3^{V - 0.05 (V-15)^2}
\end{equation}
which gives a better fit to the data (see Fig.\ 1). The relation is 
compatible with star counts predictions by the Galaxy model 
(ESA-SCI(2000)4, Torra et al.\ 1999) using Hakkila et al.\ (1997) 
extinction law. 

Average star density of ($V<17$) stars equals 15100 stars/degree$^2$ 
close to the Galactic plane ($|b| < 20^o$) and 1900 stars/degree$^2$
away from it ($|b| > 20^o$). This is somewhat larger than the 
corresponding values (6100 and 1200 stars/degree$^2$) from the 
Galaxy model (ESA-SCI (2000)4). This is probably a statistical
anomaly; symbiotic stars tend to lie close to the galactic
plane and selection effects may emphasize regions of higher
star density.  Even so only 10\%\ of directions 
close to the Galactic plane ($|b| < 20^o$) reach the density of 
40000 ($V<17$) stars per degree$^2$. 
Spectral types of field stars cluster around an early K type at the 
bright end, reaching a mid-K for the faintest targets (Figure 2).

\begin{figure}
\plotfiddle{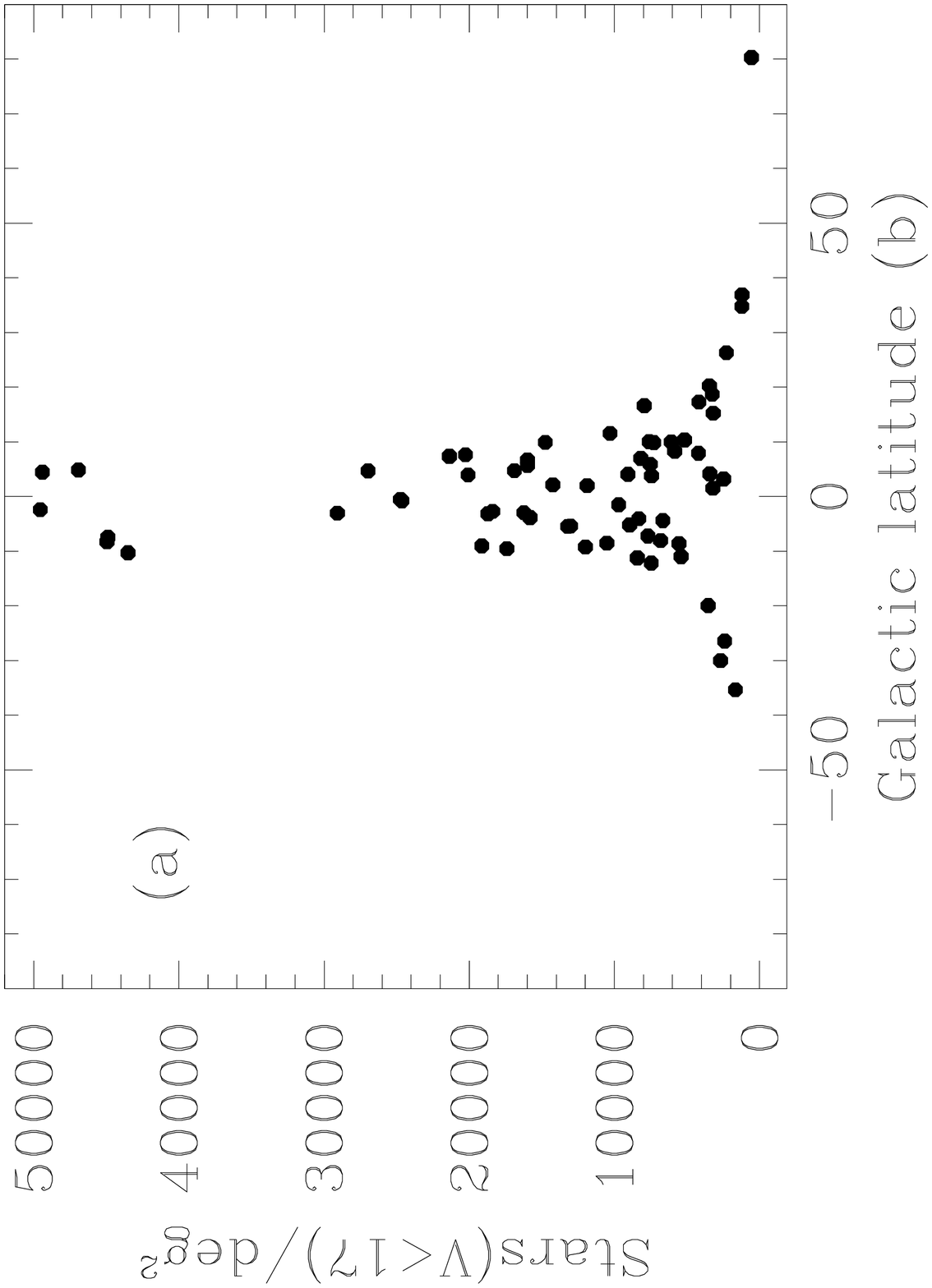}{2.6cm}{270}{25}{26}{-193}{70}
\plotfiddle{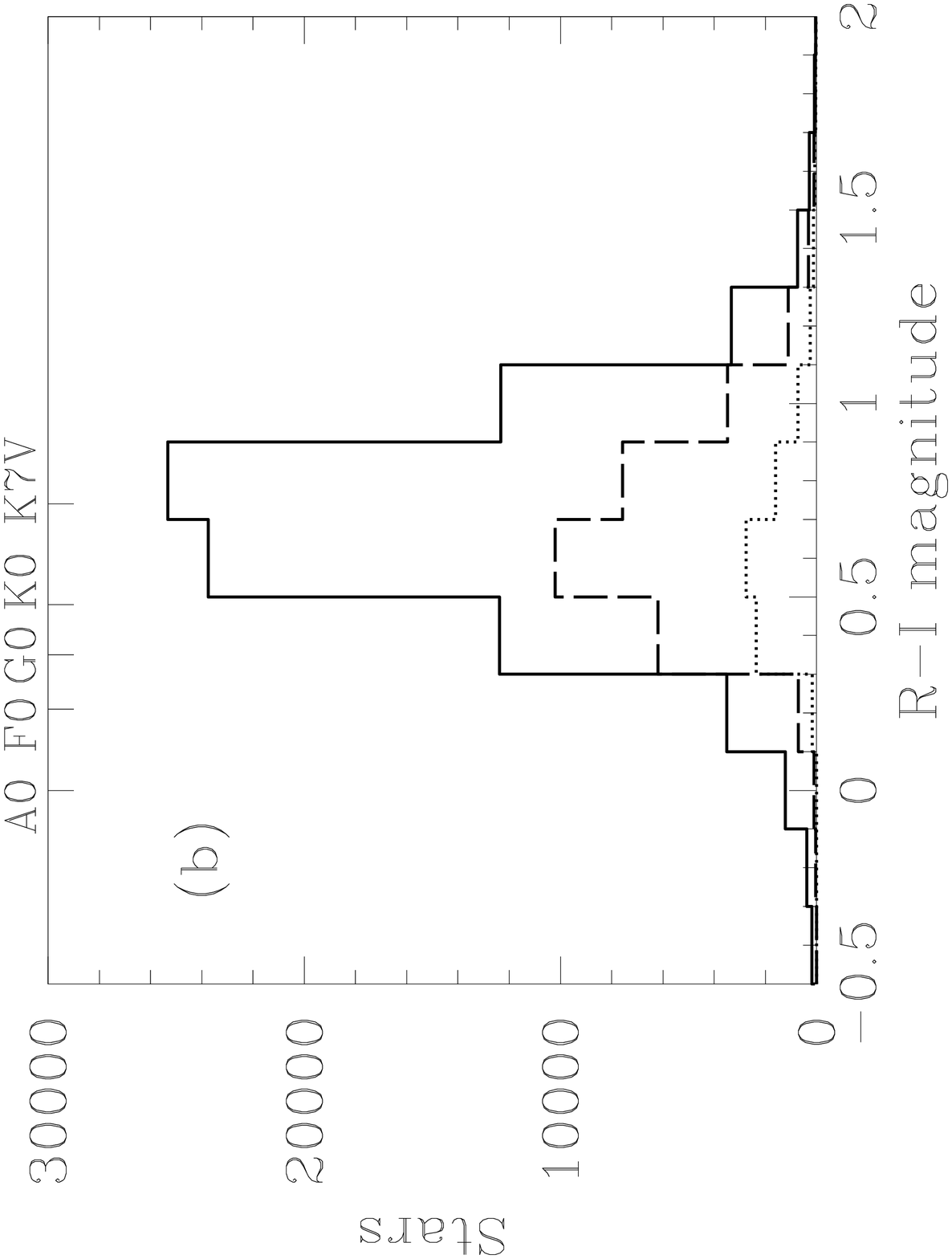}{2.6cm}{270}{25}{26}{0}{157}
\caption{(a) 
Observed star density as a function of Galactic coordinates.
(b) 
Distribution of observed colours for different magnitude 
classes: $14 < V < 16$ (dotted line), $16 < V < 18$ (dashed line), 
and $18 < V < 20$ (solid line). The latter is corrected for incompletness.
Labels on top mark colours of unreddened main sequence stars.}
\end{figure}

\section{Crowding and the sampling law}

The sampling law of the GAIA satellite guaranties that the 
arrangement of stars in the focal plane will be different for each
of the $\sim 100$ transits of a given star, providing that the 
spin and precession periods of the satellite are kept 
incommensurate. Two stars that badly overlap in one passage
have non-overlapping tracings on the next pass. This is a simple 
consequence of the fact that the length of the spectral tracing in
the dispersion direction is much larger than its width. There will 
be unfortunate cases, for example close optical doubles, with 
tracings overlapping in a significant fraction of observations. But 
such stars are rare and so of no interest to us here. Below we 
discuss the results on spectral overlaps for typical randomly 
positioned stars in the focal plane. 

Two spectra overlap with a probability $p$ if the length of the 
spectrum is larger than the free length
\begin{equation}
L = (n s)^{-1} \ln[(1-p)^{-1}]
\end{equation}
where the star density equals $n$ stars per degree$^2$ and the 
width of the stellar tracing is $s$ arc-sec ($s \sim 4.5$ arc-sec for
the Astrium design of the GAIA spectrograph). The results are 
presented in Table 1.

\begin{table}
\caption{
Severity of crowding following Eq. 2.
}
\begin{tabular}{rrrrrrr}
&&&&&&\\
\tableline
&&&&&&\\ [-5pt]
star density&average free&\multicolumn{5}{c}{probability ($p$) that the distance between}\\
$n$&         length      &\multicolumn{5}{c}{spectral heads is smaller than $L$ (arcsec)}\\
(stars/deg$^2$)&(arcsec)&$L=1$&$L=10$&$L=100$&$L=500$&$L=1000$\\ 
[4pt]
\tableline
&&&&&&\\ [-5pt]
   600&4800&0.000 &0.002 &0.021 &0.099 &0.188\\
  1200&2400&0.000 &0.004 &0.041 &0.188 &0.341\\
  3000& 960&0.001 &0.010 &0.099 &0.406 &0.647\\
  6000& 480&0.002 &0.021 &0.188 &0.647 &0.875\\
 20000& 144&0.007 &0.067 &0.501 &0.969 &0.999\\
 50000&  57&0.017 &0.159 &0.824 &1.000 &1.000\\
100000&  28&0.034 &0.293 &0.969 &1.000 &1.000\\
 [4pt]
\tableline
\end{tabular}
\end{table}

\section{Conclusions}

We shall comment on the result for a resolution of 10,000 assuming a 
spectral width of 4.5 arcsec. Results for other resolutions and spectral 
widths can be easily judged from Table 1. 

The wavelength interval of GAIA spectra covers 250~\AA\ around 
$\lambda_c = 8615 $~\AA. The spectral length for $R=10,000$ is 580 pixels, 
assuming 2 pixels per resolution element. Since each pixel covers 
1~arcsec on the sky in the dispersion direction the spectral overlap with 
another ($V<17$) star will occur if the distance between spectral heads 
is smaller than 580 arcsec. One can see from Table 1 that this happens in 
some 21 out of 100 spectra if the star density is 1200 stars/deg$^2$,  
a typical value for field stars with $V<17$ at high Galactic latitudes. 
Density of fainter stars at the same position is much higher (see Eq.\ 1), 
so overlaps with $V \sim 18$ stars are common even at high latitudes.
If one considers overlapping stars down to $V=20$ the probability for 
an overlap increases to 97\%. If the resolution were smaller 
than $R=10.000$ the length of the spectral tracing covering the 
GAIA spectral wavelength interval would be shorter. 
Still it is clear that lower spectral resolutions cannot make spectra 
free from spectral overlaps of faint stars even at high Galactic latitudes.
Fraction of spectra overlapped with $V<17$ background stars increases 
to 70\%\ at the star density of 6000 ($V<17$) stars/deg$^2$ (i.e.\ a value 
typical close to the Galactic plane).  

Spectral overlap is never complete. Even at extreme star densities of 
50.000 stars/deg$^2$ only 16\%\ of the spectra would have an overlapping 
($V<17$) spectrum starting within 10 arcsec ($= 10$ pixels) behind its head. 

Spectral tracings in the focal planes of the GAIA spectrograph will overlap 
at all star densities and at all resolutions. The policy cannot and should 
not be to keep only the spectra that are free from overlaps. Should this 
be the case one would throw away most of the collected GAIA spectra.

\end{document}